\documentstyle [12pt,epsf]{article}
 
\newcommand{\beq}{\begin{equation}}
\newcommand{\eeq}{\end{equation}}
\newcommand{\M}{M_{_{BS}}}
\def\tt{\tilde{t}}

\begin{document}
 
\baselineskip 7.5 mm
 
\begin{flushright}
\begin{tabular}{l}
CERN-TH/98-155 \\
hep-ph/9805379
\end{tabular}
\end{flushright}
 
\vspace{12mm}
\begin{center}
 
{\Large \bf
A Strongly-Interacting Phase of the Minimal Supersymmetric Model
}
\vspace{18mm}
 
{\large
Gian F. Giudice\footnote{On leave of absence from INFN, Sezione di 
Padova, Italy.}
and Alexander Kusenko\footnote{E-mail address:
Alexander.Kusenko@cern.ch.} 
}
 
\vspace{6mm}
Theory Division, CERN, CH-1211 Geneva 23, Switzerland 

\vspace{12mm}

{\bf Abstract}
\end{center}

We argue that in the minimal supersymmetric extension of the Standard Model
with a large trilinear coupling both the fundamental Higgs boson and a bound
state of squarks (formed via strong scalar interaction) can have a
non-zero VEV.  This alters drastically the pattern of electroweak-symmetry
breaking and the Higgs phenomenology.   In particular, the upper 
bound on the supersymmetric Higgs-boson mass may be relaxed. Also, the
Higgs boson can be produced at hadron colliders through a direct coupling
to gluons. 

\vspace{8mm}
 
\begin{flushleft}
\begin{tabular}{l}
CERN-TH/98-155 \\
May 1998
\end{tabular}
\end{flushleft}

\vfill 
\pagestyle{empty}
 
\pagebreak
 
\pagestyle{plain}
\pagenumbering{arabic}

\section{Introduction}

Supersymmetry predicts the existence of numerous scalar fields with
multifarious interactions.  In particular, the two Higgs doublets of the
minimal model with softly-broken supersymmetry 
can couple to squark and slepton bilinears.  These cubic
interactions,  attractive in nature, can give rise to a spectrum of bound 
states and resonances that one might hope to discover in the
future~\cite{kkt}.   

In this paper we consider the possibility that a strong trilinear
interaction produces bound states of two squarks of the third
generation.  If tightly bound, such 
states appear as additional (composite) Higgs bosons in the low-energy
effective theory.  We will explore the empirical implications of this
unusual phase of the minimal supersymmetric standard model (MSSM).

Unlike all other coupling constants in the MSSM interaction Lagrangian,
which are restricted by
supersymmetry, the dimensionful trilinear scalar couplings arise
from supersymmetry breaking.
Although calculable in some particular models, these couplings
are largely unconstrained from both the theoretical and the empirical
standpoints.  For the lightest generations of fermions, the approximate
chiral symmetry constrains the size of the trilinear terms of the 
corresponding superpartners.  
Indeed, the trilinear couplings contribute to quark masses at the one-loop
level, and the observed small masses of the first two generations require
correspondingly small $A$ terms, barring the possibility of unnatural
cancellations.

This argument does not apply to the third 
generation, since the large fermion masses give a sizable breaking of
the corresponding chiral symmetry.  However, some important upper bounds 
on the trilinear couplings come from considering the color- and
charge-breaking vacua that may have an energy density lower than that of
the standard vacuum.  If one 
requires that the standard vacuum be the global minimum of the potential,
one obtains an upper limit on the trilinear terms~\cite{ccb}.  Some more
conservative and, hence, weaker bounds~\cite{kls} allow for the possibility
that the standard vacuum be a metastable false vacuum with a lifetime that
exceeds the age of the Universe.  The analyses of refs.~\cite{ccb,kls} give 
an upper bound on the trilinear coupling $A_t$ of the stop sector
\begin{equation}
(A_t/y_t)^2 +3 \mu^2 < \beta (m_{\tt_{L}}^2+m_{\tt_{R}}^2),
\label{Alimit}
\end{equation}
where $m_{\tt_{L}}$ and $m_{\tt_{R}}$ are the diagonal entries of the stop
mass matrix, $y_t \approx 1$ is the top Yukawa coupling, and $\beta \approx
7.5$~\cite{kls} or 3.0~\cite{ccb}, depending on whether one requires the
color- and charge-conserving vacuum to be the global minimum of the
potential.  

In section~2 we will argue that the bound (\ref{Alimit}) may not apply when
the large $A_t$ term leads to the appearance of a tight bound state.  
We will show that, whenever the trilinear coupling is very large, 
a tightly bound state with the quantum numbers of the Higgs and a binding
energy of the order of the electroweak scale can form.  Moreover, we will
argue that this state can develop a non-zero vacuum expectation value
(VEV), mix with the fundamental Higgs bosons, and alter significantly the
pattern of the electroweak symmetry breaking and the ordinary Higgs
phenomenology. 

This offers the possibility for a new realization of softly-broken
supersymmetry  at the electroweak energy scale. Here and below, we assume the
usual MSSM particle content and interaction Lagrangian.  
However, we will argue that, in the large $A_t$ regime,
the theory becomes strongly-interacting, and it chooses an unconventional
phase, while its usual ultraviolet behavior is preserved.  Appealing
features of supersymmetry, such as  
the natural gauge hierarchy, the gauge coupling unification, and the prospects
for a quantum theory of gravity, are rooted in the ultraviolet regime and are
unchanged by the soft supersymmetry-breaking trilinear couplings, even
if these are as large as several TeV.  At the same time, the electroweak 
physics may be changed drastically by the appearance of a composite Higgs
boson through stop condensation, and the predicted Higgs phenomenology has
many novel features.  

The first important consequence is that the upper limit on the mass of the
lightest Higgs boson in the MSSM may be relaxed. 
The second one is that 
the Higgs production mechanisms and decay modes are modified
by the mixing between the stop bound state and the fundamental Higgs.
Both conclusions have significant implications for the experimental search
of the Higgs boson.

\section{The Minimal Supersymmetric Model in the Large $A_t$ Regime}

Let us start by considering the trilinear coupling $A_t$
between the left-handed $Q_{_L}$ and the
right-handed $\bar U_{_R}$ third-generation squarks
and the corresponding Higgs boson $H_2$
\beq
V= A_t~
H_2  Q_{_L} \bar U_{_R}+ {\rm h.c.}
\label{tril}
\eeq
For simplicity we assume that $A_t$ is the only large trilinear coupling,
but our discussion can be generalized to more complicated cases. 

The left- and right-handed stops are not mass eigenstates.  The
mass-squared matrix has the form: 
\beq
M^2= \left ( \begin{array}{cc}
\tilde{m}_{_L}^2+m_t^2 & A_t\langle H_2\rangle - \mu m_t/\tan \beta \\ 
 & \\
A_t\langle H_2\rangle - \mu m_t/\tan \beta & \tilde{m}_{_R}^2+m_t^2
\end{array}  \right ).
\label{M}
\eeq
Here $m_t$ is the top-quark mass, $\tilde{m}_{_{L,R}}^2$ are the
supersymmetry-breaking stop-mass parameters, $\mu$ is the Higgs mixing
mass, and $\tan\beta$ is the ratio of Higgs VEVs. We define the stop
mixing angle $\theta_t$, such that $\tt_1=\cos\theta_t \tt_L
+\sin\theta_t \tt_R$ is the heavier of the two mass eigenstates, and
$\tt_2=-\sin\theta_t \tt_L +\cos\theta_t \tt_R$ is the lighter one.

Since the interaction in eq.~(\ref{tril}) provides an attractive
force mediated by the exchange of the scalar quanta, we expect that bound
states can form if the coupling $A_t$ is 
large enough, and if the exchanged particle is sufficiently light. We
consider the case in which the Higgs-boson mass is significantly lower than 
the stop masses, and we argue that the theory predicts two-stop composite
states with binding energies of the order of the electroweak scale.

This speculation is supported by the observation that our theory, in the
limit in which we neglect
all interactions other than the one in eq.~(\ref{tril}), 
is identical with the Wick-Cutkosky model~\cite{wc} (for reviews, see 
ref.~\cite{iz}). 
The Bethe-Salpeter equation for the bound state in this model,  
\beq
\left [ \left (\frac{1}{2} \M + p \right)^2 +m^2 \right ]
\left [ \left (\frac{1}{2} \M - p \right)^2 + m^2 \right ]
\psi (p) = \frac{4 i A^2}{(2\pi)^4}
\int d^4k  \frac{\psi(k)}{(p-k)^2+m_{_H}^2}, 
\label{BS}
\eeq
can be solved
analytically in the ladder approximation and in the limit of
a massless-particle exchange ($m_H=0$).
Here $M_{BS},\ m$, and $m_H$ are, respectively, the masses of the bound
state, of the component particle, and of the exchanged particle; and $A$ is
a generic trilinear coupling.
Using the known solution of eq.~(\ref{BS})~\cite{wc,iz}, we obtain that
the bound states of two lighter stops $\tt_2$ have masses
\beq
M_{S_n}= 2m_{\tt_2}~\sqrt{1-\frac{\lambda^2}{4n^2}}~~~~~n=1,2,\dots
\label{bound}
\eeq
\beq
\lambda = \frac{1}{16\pi}\left(\frac{A_t \cos\alpha \sin 2\theta_t}{\sqrt{2}
\; m_{\tt_2}}\right)^2~.
\label{lam}
\eeq
Here $\cos\alpha$ measures the $H_2$ component of the lighter Higgs scalar.

The result of eq.~(\ref{bound}) has been derived in the ladder approximation,
assuming $m_H=0$.  We note that these common approximations suffer from
some well-known drawbacks.   In particular,
the ladder approximation explicitly violates the crossing symmetry,
mistreats the relativistic limit in the case of large masses, and, in the
case of Yang-Mills theories, is not gauge-invariant.
Therefore, although one can use the solution of the Bethe-Salpeter equation
as an indication of the presence of some bound states, one should take 
eq.~(\ref{bound}) with a grain of salt.  A recently reported solution to the
Bethe-Salpeter equation in the Feshbach-Villars formulation~\cite{can2}, 
$M_{S_n}=2m_{\tt_2}~\sqrt{(1+\sqrt{1-\lambda^2/n^2})/2}$, 
gives a larger binding energy than that obtained in the ladder
approximation.  Concerning the $m_H=0$ limit,  
numerical studies~\cite{can} have shown that, in the Wick-Cutkosky
model with a massive intermediate particle, the bound state exists as long
as the coupling $\lambda$ in eq.~(\ref{lam}) is larger than some critical
value $\lambda_c$.  In addition, our case is further
complicated by the mixing of the bound state and the fundamental Higgs, as
discussed below.  

In conclusion, the analyses of the Bethe-Salpeter equation support our
speculation that there exists a region in the parameter space of the MSSM
in which two stops form a bound state with binding energy of the order 
of the electroweak scale. This occurs for values of the coupling $\lambda$
of order unity, where the theory becomes strongly interacting.  In terms of
the supersymmetry-breaking parameters, this happens for large values
of $A_t$ (say of the order of a few TeV) and considerably smaller values of
$\tt_2$ (of the order of a few hundred GeV). This mass hierarchy
can be the result of an approximate cancellation in the determinant of
the stop mass matrix in eq.~(\ref{M}), which indeed occurs in the
large $A_t$ regime.  

Although lacking a rigorous proof, this description of
the large $A_t$ phase appears quite plausible to us. 
We hope that future studies will give further support to this
conclusion. Since the relevant degrees of freedom in the problem are scalar
fields, a lattice approach appears a well-suited tool for analyzing the
dynamical properties of this theory. 

  From a phenomenological point of view, the lightest $s$-wave bound state,
denoted here by $S$, plays an important role.  We expect that $S$ is  
a color-singlet $s$-wave combination of $[\tt_2^\star \tt_2]$.
The theory also contains a series of resonances with different Standard
Model (SM) quantum
numbers, {\it e.\,g.}, the color-octet combination of $[\tt_2^\star \tt_2]$, 
and the charge-4/3 color sextuplet or antitriplet $[\tt_2 \tt_2]$.
Depending on the supersymmetric parameters, bound states involving $\tt_1$ or
the sbottom may also be formed. For instance, consider the state
$S^\prime =[\tt_2^\star \tt_1]$. Solving the corresponding
Bethe-Salpeter equation in the ladder approximation and in the limit
of massless-particle exchange, we find the analogue of eq.~(\ref{bound})
for the lowest-lying bound-state mass
\beq
M_{S^\prime}= \left[ \left( m_{\tt_1}+m_{\tt_2}\right)^2 -\frac
{(A_t \cos\alpha \cos 2\theta_t)^4}{(32\pi)^2 m_{\tt_1} m_{\tt_2} 
}\right]^{1/2}
\eeq
Likewise, electrically-charged resonances can be formed by the bound states
of the sbottom and the stop $[\tilde{b}_2^\star \tt_2]$. Unless the sbottom
and the charged Higgs (which mediates the interaction) are significantly
lighter than the mass scale of the trilinear coupling, the binding energy
of the charged resonance should be small.

At sufficiently low energies, the stop fields do not represent the appropriate
degrees of freedom in the low-energy effective theory, which must be
described in terms of the bound states. We cannot compute perturbatively
the effective potential involving the fundamental Higgs doublets and
the bound states, and therefore  we have to consider the possibility that  
the system can be in any of the following phases.  

{\bf (i) SU(2)$\times$U(1)-symmetric phase}, in which all scalars have zero
VEVs. 

{\bf (ii) Ordinary Higgs phase}, in which the two Higgs doublets have
non-zero VEVs, but both the stop fields and the $s$-wave bound states 
have zero VEVs, $\langle \tt \rangle =\langle S \rangle = 0$.  
This is the usual (color- and charge-preserving) vacuum.  

{\bf (iii) Color- and charge-breaking vacuum} with  $\langle \tt 
\rangle \neq 0$. This phase is possible if the $A_t$ coupling is
sufficiently large.  However, this is not the only possibility in the limit
of strong trilinear coupling.  In addition, there may exist a

{\bf (iv) composite Higgs phase}, in which 
the two fundamental Higgs bosons and the bound state $S$ have non-zero VEVs,
although $\langle \tt_1
\rangle =\langle \tt_2
\rangle =0$. In this case $S$ plays the role of an additional Higgs
field, contributing to the electroweak breaking.

The upper bound (\ref{Alimit}) on the value of the trilinear coupling was
derived under an implicit assumption that, as the coupling $A_t$ increases,
the system remains in phase {\bf (iii)}.  This is not necessarily the
case.  If a transition to phase {\bf (iv)} takes place at some large value
of $A_t$, then the bound (\ref{Alimit}) does not apply.  It is beyond the
scope of our analysis to determine the phase diagram of the theory.  A
calculation on a lattice is probably the only reliable tool for answering
this question, which depends essentially on the non-perturbative physics. 
Here we just assume that, in a certain range of parameters, the system 
chooses phase {\bf (iv)}, and we will investigate the theoretical and 
experimental consequences of this assumption.

It appears quite plausible that $S$ can develop a non-zero VEV when
$SU(2)\times U(1)$ is spontaneously broken by the Higgs field. Indeed, as
we will discuss in the next section, eq.~(\ref{tril}) implies
a mixing mass term between $S$ and $H_2$. A non-zero VEV for $H_2$ generates
a tadpole term that triggers a non-zero $\langle S \rangle$.

In phase {\bf (iv)}, $S$ participates in the electroweak breaking and
therefore behaves as a third Higgs boson.  Since the overall 
size of electroweak breaking is fixed by the $W$ and $Z$ masses, the
relation between the Yukawa couplings and the known fermion masses changes.  
Moreover, the top-quark mass receives new contributions from large one-loop
corrections involving the stop exchange and the $A_t$ coupling, and also
from a tree-level chirality-violating effective coupling
between the top quark and $S$, after integrating out the gluino. As a result,
the running top-quark Yukawa is smaller than that usually predicted, and
very low values of $\tan\beta$ are consistent with the absence of a Landau
pole up to very high energies.
However, the effective coupling of $S$ to quarks is restricted 
by the presence of flavour violations mediated by neutral Higgs exchange.

It is interesting to study how the coupling $\lambda$ in eq.~(\ref{lam})
evolves with the renormalization scale. The corresponding RG equation
depends on several unknown supersymmetry-breaking parameters. However,
it is easy to show that there exists a range of parameters
in which the $A_t$ coefficient grows at low energies
faster than $\tilde{m}_{_{L,R}}$, because of the gluino mass effect. Under
these conditions, the coupling $\lambda$ remains small down to some 
low-energy scale at which an approximate cancellation takes place in 
the determinant of the mass matrix in eq.~(\ref{M}).  Then
$m_{\tt_2}$ becomes significantly lower than the typical scale
of $\tilde{m}_{_{L,R}}$. This triggers a rapid growth of $\lambda$ 
to some large, non-perturbative values. 
In this way it is possible to have a situation in which the high-energy
behavior of the theory is analogous to the ordinary MSSM behavior, 
with the stop fields describing dynamical degrees of
freedom. Non-perturbative effects arise only at energies of the order of 
the electroweak scale, where at least one of the stops is confined in a
bound state. In particular, this means that the prediction of
gauge-coupling unification remains unaffected. The non-perturbative 
$A_t$ interactions give an effective contribution that can be absorbed
in the unknown low-energy threshold
corrections. On the other hand, because of the new contributions to the
top-quark mass, this theory does not reproduce the usual perturbative
results for the extrapolation of the third-generation Yukawa couplings
to high energies.

\section{Phenomenological Implications}

In the previous section we argued that there exists a parameter
range in which the $s$-wave
color-singlet bound state $S=[\tt_2^\star \tt_2]$ develops a
VEV. Under the same circumstances, we also expect that $S$ mixes with the
fundamental Higgs scalars. Let us consider the low-energy effective theory
valid at the electroweak scale. It contains the new scalar degree of
freedom, namely the composite Higgs boson $S$. 
The very coupling in eq.~(\ref{tril})
that gives rise to $S$ in the low-energy effective theory causes the mixing
between composite and fundamental Higgs bosons.
The effective potential contains a term linear in both $S$ and $H_2$: 
\beq
V_{eff}=m^2_{HS}SH_2 +{\rm h.c.}
\eeq
Here $m^2_{HS}=A_tf$, where $A_t$ is the trilinear coupling and $f$ is
a form factor that depends on the
strong dynamics and is of the order of the electroweak scale. 

Knowledge of the effective potential requires a non-perturbative calculation.
Since there is effectively only one mass scale in the problem (the electroweak
scale), it is difficult to speculate on the form of the potential. In general,
we expect some polynomial interactions in $S$ and $H_{1,2}$ with couplings (in
units of the electroweak scale) of order one. The perturbative predictions
of the Higgs mass spectrum are no longer valid. In particular, new Higgs
quartic couplings in the effective potential are generated. Therefore, one
cannot trust the usual upper bound on the supersymmetric Higgs mass, although
we are not able to quantify the non-perturbative contribution.  
Nevertheless, it is interesting that the absence of a very light Higgs 
boson does not necessarily imply the demise of low-energy supersymmetry.  
In fact, if the lightest Higgs boson is heavy, this may 
signal the strongly-coupled regime of the MSSM with stop condensation. 

Let us now turn to the phenomenology at high-energy colliders. 
In the regime under consideration, the theory predicts various bound states
and resonances 
with different SM quantum numbers, as illustrated in sect.~2. At hadron
colliders $[\tt^\star \tt ]$ states can be singly produced via gluon
fusion, while $[\tt \tt ]$ states, which carry flavour number, 
must be produced in pairs or accompanied by two top quarks. 
If it is kinematically possible, heavy resonances will decay through the
strongly-coupled scalar dynamics into a pair of lighter Higgses (fundamental
or composite). Otherwise, their decay modes are quite dependent on the
mass spectrum and the supersymmetry-breaking parameters. 

Here we want to concentrate on the lightest state in the scalar sector, $h$. 
As previously argued, this is a linear combination of the fundamental
Higgs bosons and the $s$-wave, color-singlet bound state $S$
\beq
|h\rangle = \frac{\cos \phi}{\sqrt{2}}\left( \sin \alpha |H_1^0\rangle
+\cos \alpha |H_2^0\rangle \right) + \sin \phi |S\rangle ~.
\eeq
The experimental
signatures then depend on the unknown mixing angles, and, in particular, 
on $\sin \phi$, which sets the fraction of composite Higgs in the lightest
scalar. Since the properties of the fundamental Higgs components are well
known, we focus here on the properties of the $S$ component of the scalar
state $h$.

We note in passing that the kinematics allow $S$ to form a true bound
state since its formation time is 
significantly shorter than its lifetime. Indeed, the formation time is
characterized by the inverse of the binding energy, {\it i.\,e.} the
inverse of the electroweak scale. On the other hand, the $S$ decay 
occurs via perturbative interactions, either QCD in the case of 
stop-antistop annihilation, or weak interactions in the case of stop
decay. Therefore, the 
$S$ decay width is considerably smaller than the typical electroweak scale.
This is in contrast with the case of stoponium bound states previously
considered in the literature~\cite{stop,drees}, in which perturbative QCD 
provided the attractive force, and the binding energy was only about 1$\%$
of its mass.  

The $S$ bound state can decay through annihilation of its constituents or
via decay of the constituents themselves. In the latter case, the most likely 
processes are stop decays into charginos or neutralinos, $\tt_2 \to
t \chi^0$ and $\tt_2 \to b \chi^+$, which however can occur only if
$m_h > m_{\tt_2}+m_{\chi^0}+m_t$ and $m_h > m_{\tt_2}+m_{\chi^+}+m_b$,
respectively. The process $\tt_2 \to c \chi^0$
is strongly suppressed by flavor conservation and the decay $\tt_2 \to
t \tilde{g}$ is probably inhibited by the large gluino mass. 
For instance the decay rate into charginos $\chi^+_i$ ($i=1,2$) 
is\footnote{Complete
formulae for stop and stoponium decay rates can be found in ref.~\cite{drees}.}
\beq
\Gamma (\tt_2 \to b \chi^+_i ) = m_{\tt_2}~\frac{g^2}{16\pi}
\left( V_{i1} \sin \theta_t +\frac{m_t V_{i2}}{\sqrt{2} m_W \sin\beta}
\cos \theta_t \right)^2 \left( 1-\frac{m_{\chi^+_i}^2}{m_{\tt_2}^2}
\right)^2 ~.
\label{chargi}
\eeq
Here $V$ is the usual chargino diagonalization matrix and $\Gamma (S)=
2\Gamma (\tt_2 )$.

The stop annihilation process can lead to a pair of gauge bosons, of
charginos, of neutralinos, or to a quark-antiquark pair. The annihilation
into gluons is likely to dominate, with a rate given by
\beq
\Gamma (S \to gg) = \frac{8}{3} \alpha_s^2 \Delta ~.
\label{gg}
\eeq
In the non-relativistic limit, $\Delta$ is given in terms of the 
$S$-wavefunction at the origin $R(0)$ by the expression $\Delta
=|R(0)|^2/M_S^2$. In our case, the bound state is relativistic, as the
binding energy is of the order of the particle masses. Nevertheless, we
can still parametrize the decay rate into gluons by eq.~(\ref{gg}), taking
$\Delta$ as an unknown parameter of the order of a few hundred GeV.

Because of the QCD charge of its constituents, the $S$ component of $h$
has a direct coupling to gluons, which can significantly increase the
Higgs production rate at hadron colliders. In the narrow-width
approximation, the production cross section is
\beq
\sigma (pp\to h+X) = \frac{\pi^2}{8 m_h^3}\sin^2 \phi ~\Gamma (S\to gg)
\int_\tau^1 \frac{dx}{x}~\tau G(x,Q^2)G(\tau /x,Q^2) ~,
\eeq
where $\tau=m_h^2/s$, $\Gamma (S\to gg)$ is given in eq.~(\ref{gg}) and $G$
describes the gluon density. The cross
section at $p\bar p$ colliders is the same, since the production occurs
via gluon fusion. Although the $S$ width is typically several GeV, the
narrow-width approximation is not necessarily inadequate since the
experimental dijet mass resolution is about 10{\%}. The numerical values of
the total cross sections (in units of $(\Delta/100\,{\rm GeV}) \sin^2\phi$)
for the Tevatron and the LHC are shown in fig.~1, using $Q^2=m_h^2$.

\begin{figure}
\setlength{\epsfxsize}{3.5in}
\centerline{\epsfbox{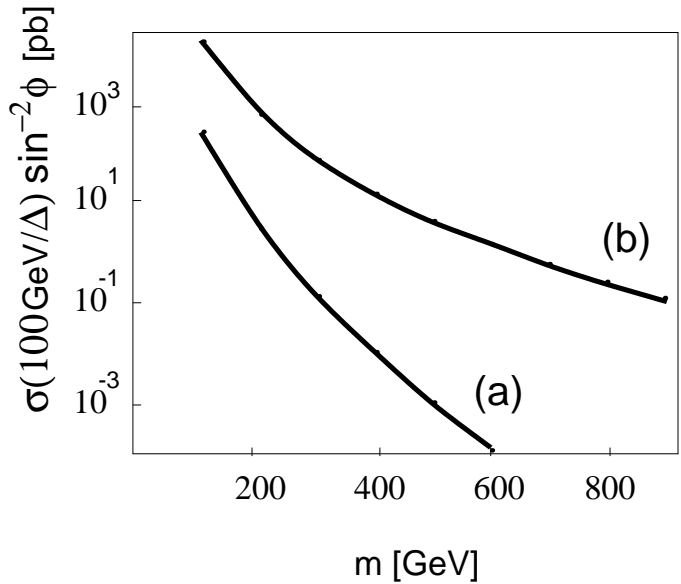}}
\caption{ Production cross-section for the scalar state {$h$} in units
of {$\Delta (\sin^2 \phi)\; /100$}~GeV, as a function of the {$h$} mass.
Curve (a) refers to the Tevatron ({$\sqrt{s}=1.8$}~TeV) and curve (b)
to the LHC ({$\sqrt{s}=14$}~TeV).
}
\label{fig1}
\end{figure}

The experimental signature of the composite Higgs depends on the decay
modes.
If $\Delta \simeq M_S$, then the two-gluon decay
rate in eq.~(\ref{gg}) is larger than
the one in eq.~(\ref{chargi}). Nevertheless, depending on the 
supersymmetry-breaking terms
and the non-perturbative parameter $\Delta$, it is possible that the 
mode in eq.~(\ref{chargi}) or other decay modes into supersymmetric
particles become the dominant channel. An experimental search should
consider these various possibilities. 

When the decay into two gluons dominates, the composite Higgs appears as
a peak in the invariant-mass distribution of the dijet cross section.
CDF has searched for anomalous resonances in the dijet system
and derived limits on the corresponding
production cross-section~\cite{CDF}. Unfortunately, the present limits are
too weak to set meaningful bounds on particles produced from gluon
fusion, such as the composite Higgs $S$. A more promising way of searching
for  $S$ is to employ its decay into two photons, since the background can be
effectively reduced by simple acceptance cuts. Assuming that the 
non-perturbative form factor $\Delta$ is the same in the decay into
gluons and photons, we find
\beq
\Gamma (S\to \gamma \gamma ) = \frac{8}{9} \left( \frac{\alpha}{\alpha_s}
\right)^2 \Gamma (S\to gg) ~.
\eeq
This small branching ratio into two photons can be used as a distinguishing
feature of the composite Higgs both at the Tevatron and at the LHC. 
When the decay into supersymmetric particles 
dominate, the experimental signature is quite dependent on the model
parameters, but in general it leads to a structure of decay chains with many
particles in the final state. 
Leptonic decays of charginos and heavy neutralinos provide the
best channel for discovery. 

The production cross-section for $S$ at the Tevatron is limited by the
small gluon luminosity at $\sqrt{s}=1.8$~TeV.  Heavier resonances produced
in quark-antiquark collisions could then provide the discovery mode.  A
good candidate is the color-octet $p$-wave bound state
$P_8=[\tt_2^*\tt_2]$. Since $P_8$ is likely to have binding energy
comparable to its mass,
the $p$-wave suppression may not be important\footnote{For a calculation of
  the $p$-wave suppression of stoponium production in the non-relativistic
  limit, see ref.~\cite{bigi}.} and its production cross-section at the
Tevatron can be significant.  Depending on the masss spectrum, $P_8$ can 
decay into lighter bound states, as in $P_8 \to h g$ or $P_8\to hS_8$,
where $S_8$ is the $s$-wave color-singlet state.  Alternatively, it can
disintegrate through stop decay.  In any case, the final state presents a
rich hadronic activity.

\section{Conclusions}

We have argued here that the well-known and well-studied MSSM may manifest
itself in a very different way than usually expected. In a possible range
of the yet unknown supersymmetry-breaking parameters, the theory becomes
strongly interacting. The scalar partners of the top quarks form bound states
that can mix with the fundamental Higgs bosons.  It appears  
plausible that the color-singlet $s$-wave bound state of two stops develops
a non-zero VEV and takes part in the electroweak symmetry breaking. The 
Higgs phenomenology of the MSSM can differ drastically from the predictions
based on perturbative analyses. In particular, the upper bound on the
lightest Higgs boson mass may be relaxed.  Moreover, the Higgs boson can be
produced at a hadron collider via its bound-state component, which has a
direct coupling to two gluons. 

This strongly-interacting regime occurs when the trilinear coupling $A_t$
becomes large, of the order of a few TeV, and drives the stop mass to values
of a few hundred GeV, below the typical scale of the
stop supersymmetry-breaking masses. 
The effective scalar potential at the electroweak scale is dominated by
non-perturbative effects, and, therefore, the usual analysis of the color- 
and charge-breaking minima does not apply. It is quite plausible that the
trilinear scalar interaction is weak at high energies and enters its 
non-perturbative regime only near the electroweak scale.  The
usual high-energy behavior of the MSSM remains unchanged and, in particular,
the predictions for the gauge-coupling unification are unaffected.

We emphasize that the stop condensation and the composite scalar states we 
hypothesized are not meant to
replace the Higgs sector. They just appear as a dynamical consequence of
the ordinary MSSM in the regime of large $A_t$. Fundamental
Higgs bosons are present in the theory, along with the composite states bound
by the strong scalar force.  In this respect, our approach
is very different from the idea of top-condensate models~\cite{top}, in 
which new effective interactions eliminate the need for a fundamental
Higgs. However, even when the effective interaction is mediated by a
fundamental field~\cite{ross}, one is forced 
to introduce a large mass scale, which brings back the naturalness problem
of the SM Higgs sector. In our case, it is the ultraviolet behavior of
softly-broken supersymmetry that is ultimately responsible for protecting
the gauge hierarchy. 

We have presented a plausible scenario for the dynamical behavior of MSSM
in the large $A_t$ regime.  However, only a truly non-perturbative
calculation  could prove our speculations and quantify our predictions. 
We hope that lattice simulations of the theory (which essentially involves
only scalar degrees of freedom) will allow one to make further progress in
understanding the strong-coupling phase of the MSSM. 

\bigskip

We wish to thank G.~Altarelli, R.~Barbieri, F. Feruglio, M.~Mangano,
R.~Rattazzi, M.~Shaposhnikov, and C.~Wagner for very helpful discussions.


\begin{thebibliography}{99}

\bibitem{kkt} A. Kusenko, V.A.~Kuzmin, and I.I.~Tkachev, Phys. Lett. B,
  in press (hep-ph/9801405).   


\bibitem{ccb} M.~Claudson, L.~Hall, and I.~Hinchliffe, Nucl. Phys. B228
 (1983) 501; L.~\'Alvarez-Gaum\'e, J.~Polchinski, and M.~Wise,
  Nucl. Phys. B221 (1983) 495; J.M.~Fr\`ere, D.R.T.~Jones, and S.~Raby,
  Nucl. Phys. B222 (1983) 11; C.~Kounnas, A.B.~Lahanas, D.V.~Nanopoulos,
  and M.~Quir\'os, Nucl. Phys. B228 (1983) 501; J.P.~Derendinger and
  C.A.~Savoy, Nucl. Phys. B237 (1984) 307; M.~Drees, M.~Gl\"uck, and 
  K.~Grassie, Phys. Lett. B157 (1985) 164; J.F.~Gunion, H.E.~Haber, and
  M.~Sher, Nucl. Phys. B306 (1988) 1; H.~Komatsu, Phys. Lett. B215 (1988)
  323;  P.~Langacker and N.~Polonsky, Phys. Rev. D50 (1994) 2199;
  A.J.~Bordner (hep-ph/9506409); J.A.~Casas, A.~Lleyda, and C.~Mu\~noz,
  Nucl. Phys. B471 (1996)~3;  S.A.~Abel and C.A.~Savoy (hep-ph/9803218). 

\bibitem{kls} A.~Kusenko, P.~Langacker, and G.~Segr\`e, Phys. Rev. D 54
  (1996) 5824; A.~Kusenko and P.~Langacker, Phys. Lett. B 391 (1997) 29.

\bibitem{wc} G.C.~Wick, Phys. Rev. 96 (1954) 1124; R.E.~Cutkosky,
  Phys. Rev. 96 (1954) 1135.  

\bibitem{iz} C.~Itzykson and J.-B.~Zuber, {\it Quantum Field Theory},
  McGraw-Hill, New York 1980; N.~Nakanishi, Prog. Theor. Phys. Suppl.
95 (1988) 1; Z.K.~Silagadze (hep-ph/9803307).

\bibitem{can2} M.~Barham and J.~Darewych, J.~Phys.~A 31 (1998) 3481.

\bibitem{can} E.~Zur~Linden and H.~Mitter, Nuovo Cim. B61 (1969) 389;
C.-R.~Ji and R.J.~Furnstahl, Phys. Lett. B167 (1986) 11; 
L.~Di~Leo and J.~Darewych, Can. J. Phys. 70 (1992) 412; T.~Nieuwenhuis
and J.A.~Tjon, Phys. Rev. Lett. 77 (1996) 814.

\bibitem{stop} V. Barger and W.-Y. Keung, Phys. Lett. B211 (1988) 355;
K.~Hagiwara, K.~Kato, A.D.~Martin, and C.-K.~Ng, Nucl. Phys. B344 (1990) 1; 
H.~Inazawa and T.~Morii, Phys. Rev. Lett. 70 (1993) 2992. 

\bibitem{drees} M.~Drees and M.M.~Nojiri, Phys. Rev. D49 (1994) 4595.

\bibitem{CDF} F. Abe {\it et al.} (CDF Coll.), Phys. Rev. D55 (1997) 5263.

\bibitem{bigi} I.I.~Bigi, V.S.~Fadin, and V.~Khoze, Nucl. Phys. B377 (1992)
  461.

\bibitem{top} Y. Nambu, {\it in} Proc. Int. Workshop on New Trends in
Strong Coupling Gauge Theories, Nagoya, 1988, eds. M. Bando, T.~Muta,
and Y.~Yamawaki; W.A.~Bardeen, C.T.~Hill, and M.~Lindner, Phys. Rev. D41
(1990) 1647.

\bibitem{ross} D.E.~Clague and G.G.~Ross, Nucl. Phys. B364 (1991) 43.

\end{thebibliography}
\end{document}